\documentstyle[aps,pre]{revtex}

\begin{document}
\draft
\title{Soliton tunneling with sub-barrier kinetic energies}
\author{J. A. Gonz\'alez}
\address{Centro de F\'\i sica, Instituto Venezolano de Investigaciones \\
Cient\'\i ficas, Apartado Postal 21827, Caracas 1020-A, Venezuela}
\author{A. Bellor\'{\i}n}
\address{Departamento de F\'\i sica, Facultad de Ciencias, Universidad Central de\\
Venezuela, Apartado Postal 47588, Caracas 1041-A, Venezuela}
\author{L. E. Guerrero}
\address{Departamento de F\'\i sica, Universidad Sim\'on Bol\'\i var, Apartado 89000,%
\\
Caracas 1080-A, Venezuela}
\date{\today }
\maketitle

\begin{abstract}
We investigate (theoretically and numerically) the dynamics of a soliton
moving in an asymmetrical potential well with a finite barrier. For large
values of the width of the well, the width of the barrier and/or the height
of the barrier, the soliton behaves classically. On the other hand, we
obtain the conditions for the existence of soliton tunneling with
sub-barrier kinetic energies. We apply these results to the study of
soliton propagation in disordered systems.
\end{abstract}


\preprint{HEP/123-qed}

\ 

\narrowtext

The escaping process of a particle from a potential well, as that shown in
Fig. 1 (barrier crossing), is a problem of great importance in almost all
areas of physics \cite{Hanggi}.

In the case of a classical particle, this escape should occur over the
barrier with the help of external perturbations (e.g., thermally activated
barrier crossing) \cite{Hanggi}. On the other hand, a quantum particle can
perform tunneling with certain probability $p<1$.

In the present letter we address the question: {\it what happens if, in the
potential well, instead of a point-like particle we have a soliton?} This is
very relevant to Skyrmion models for nucleon physics, the motion of fluxons
in long Josephson junctions with impurities, the dynamics of domain walls in
ferroelectric materials in the presence of inhomogeneous electric fields,
and many other physical systems where the solitons move in a potential
created by inhomogeneities and external forces \cite
{Bishop,Skyrme,Adkins,Scott}.

It is well-known that a soliton can behave as a classical particle in some
physical systems \cite{Bishop,Kivshar1}. However, recently there has been a
great interest in non-classical behaviors of the soliton \cite
{Kivshar2,Gonzalez1,Gonzalez2,Gonzalez3,Kivshar3,Campbell,Guerrero2,Kalbermann2}. In
particular, we are interested in extremely surprising phenomena that can
occur when the soliton behaves as an extended object \cite
{Gonzalez1,Gonzalez2,Gonzalez3}.

Among these phenomena is the soliton tunneling suggested by K\"albermann in
a beautiful paper \cite{Kalbermann}. However, this was a numerical work and
the tunneling with sub-barrier kinetic energies was not observed in the
specific situations in which the numerical experiments were performed.

In the present letter we show (theoretically and numerically) that the
tunneling with sub-barrier kinetic energies is, indeed, possible!

As an example we consider the perturbed $\phi ^4$-equation:

\begin{equation}
\phi _{tt}-\phi _{xx}-\frac 12(\phi -\phi ^3)=F(x).  \label{Eq1}
\end{equation}

The external force $F(x)$ is such that a point-like soliton would feel an
effective potential like that shown in Fig. 1.

When the soliton is treated as a point-like particle, the zeroes of $F(x)$
are equilibrium points \cite{Gonzalez2}. The zeroes $x_0$ ($F(x_0)=0$) for
which $\left[ \frac{dF(x)}{dx}\right] _{x=x_0}>0$ are stable equilibrium
positions. In the opposite case, they are unstable.

For our theoretical calculations we will use the force $F(x)$ defined in the
following way: 
\begin{equation}
F(x)=F_1(x)\text{, for }x<x^{*},  \label{Eq2}
\end{equation}
\begin{equation}
F(x)=c\text{, for }x>x^{*},  \label{Eq3}
\end{equation}
where $F_1(x)=\frac 12A(A^2-1)\tanh (Bx)+\frac 12A(4B^2-A^2)\frac{\sinh (Bx)%
}{\cosh ^3(Bx)}$, $x^{*}$ $\left( x^{*}>0\right) $ is the point where $%
F_1(x) $ has a local minimum $\left( \frac{dF(x^{*})}{dx}=0\right) $, and $%
c=F_1(x^{*})$.

The condition $\left| F(-\infty )\right| =\frac 12A\left| A^2-1\right| <%
\frac 1{3\sqrt{3}}$ should hold for the stability of the soliton as a whole.
This force allows us to solve the problem of soliton dynamics in a
neighborhood of the equilibrium points \cite{Gonzalez1,Gonzalez2,Gonzalez3}.
For instance, the stability problem $\left( \phi (x,t)=\phi
_k(x)+f(x)e^{\lambda t}\right) $ of the equilibrium point $x_0=0$ is reduced
to the eigenvalue problem $\hat Lf=\Gamma f$, where $\hat L=-\partial
_x^2+\left( \frac 32A^2-\frac 12-\frac{3A^2}{2\cosh ^2(Bx)}\right) $ and $%
\Gamma =-\lambda ^2$. The eigenvalues of the discrete spectrum are given by $%
\Gamma _n=-\frac 12+B^2(\Lambda +2\Lambda n-n^2)$ where $\Lambda (\Lambda
+1)=\frac{3A^2}{2B^2}$.

Our analysis reveals that if $A^2>1$ and $4B^2<1$, the force given by Eqs. (%
\ref{Eq2})-(\ref{Eq3}) possesses the desired properties, i.e. there is a
zero that would correspond to a stable equilibrium position in a point $x=-d$
($d>0$) and a zero in the point $x=0$ that would correspond to an unstable
equilibrium position and would serve as a potential barrier. For $x>0$ the
potential is a monotonically decreasing function.

In fact, if $2B^2\left( 3A^2-1\right) <1$, then the soliton behaves
classically. In this case, the soliton {\it feels} the barrier in the point $%
x=0$. If the soliton is situated in a vicinity of point $x=0$ with zero
initial velocity and with the center of mass in a point $x<0$, it will not
move to the right of point $x=0$.

On the other hand, if $2B^2\left( 3A^2-1\right) >1$, the soliton will move
to the right, crossing the barrier even if its center of mass is placed in
the minimum of the potential and its initial velocity is zero (see Fig. 2).
In this case the soliton performs tunneling with sub-barrier kinetic energy!

We should remark that this phenomenon is possible only when the distance $d$
between the minimum of the potential well and the maximum of the potential
barrier holds the inequality $d<2.17$, where $d=\left( \frac 1B\right) 
\mathop{\rm arccosh}
\left( \sqrt{\frac{A^2-4B^2}{A^2-1}}\right) $. This can be interpreted in
the sense that the {\it wavelength} of the soliton should be comparable with
the width of the potential well.

In this context, the work \cite{Kalbermann2} addresses the differences and
similarities of soliton phenomena with those of point particles. Although
the studied phenomena are very different, we can say that the main conclusion 
is in consistency with our previous 
papers \cite{Gonzalez1,Gonzalez2,Gonzalez3} and the present one:
the soliton behaves as a particle only when the width of the potential
makes the soliton appears as point-like. Otherwise, the soliton can have
wave-like extended character.

The condition for soliton tunneling can be written in more physical terms.
In fact, the force $F(x)$ can be defined by the value $F_0=\left|
\lim_{x\rightarrow -\infty }F(x)\right| $, the width of the barrier $S$ and
the local maximum $F_m$ of the force between the points $x=-d$ and $x=0$
(see inset in Fig. 1). In these terms, the approximate condition for the
existence of soliton tunneling is $F_0>\frac{F_mS^2}6$. This inequality
shows that greater values of $F_0$ support the tunneling, while greater
values of $F_m$ and $S$ can thwart the tunneling.

We have performed numerical experiments with the force $F(x)$ defined as in
Eqs. (\ref{Eq2})-(\ref{Eq3}) and with many other functions which produce an
effective potential as that shown in Fig. 1. We have been able to control
the values of $F_0$, $F_m$ and $S$. Figure 3 shows the numerical
experiments. The dots on the curve separates two zones: the zone in which
the soliton tunneling is possible (upper zone) and the zone in which the
soliton tunneling is impossible (lower zone). Note that the relation $F_0=%
\frac{F_mS^2}6$ is approximately satisfied. With other forces, the results
are qualitatively equivalent.

Sometimes, it is convenient to see the condition for the existence of
soliton tunneling in terms of a parameter that defines the potential. An
important characteristic of the potential is the height of the potential
barrier: $V_m$. Figure 4 shows the relationship between $F_0$ and $V_m$
while the bifurcation condition $2B^2\left( 3A^2-1\right) =1$ holds. For
points above the curve, the tunneling occurs. For points under the curve,
the soliton behaves classically. That is, as was expected, the height of the
potential barrier is an opposing factor with respect to the tunneling.

Note that the values of the potential $V(x)$ (or the force $F(x)$) for $x\ll
-d$ can influence the tunneling. This is in contrast with the behavior of a
point-like classical particle.

We should remark directly that, in this phenomenon, there is no violation of
the energy conservation law. In fact, the energy of the whole system

\begin{eqnarray}
E\equiv \int\nolimits_{-\infty }^\infty \left[ \frac 12\left( \frac{\partial
\phi }{\partial t}\right) ^2+\frac 12\left( \frac{\partial \phi }{\partial x}%
\right) ^2-\frac 14\phi ^2+\frac 18\phi ^4 \right. \nonumber \\
 \left. -F(x)\phi +c\right] dx  \label{Eq5}
\end{eqnarray} 
is conserved. In this case, the soliton does not behave as a classical
point-like particle. The soliton, as an extended object, possesses a
wave-mechanical behavior. Even when the center of mass is situated in a
point of minimal potential energy and with zero kinetic energy, the system
as a whole can have enough energy to make the soliton able to cross the
barrier (although the center of mass goes through the barrier).
K\"albermann's example of a high jumper is felicitous \cite{Kalbermann}.
However, we should notice the presence of nonlocal effects during this
process.

The propagation of solitons in disordered media has been studied intensively
in last years \cite{Gredeskul}. There is consensus in the conclusion that
nonlinearity can modify the effects of localization and the transmission is
improved. Nevertheless, even in a nonlinear system supporting solitons, if
the latter behave as point-like particles, they can be trapped in the zeroes
of the force $F(x)$. We stress that the phenomenon of soliton tunneling can
enhance even more the transmission.

Consider Eq. (\ref{Eq1}) with $F(x)$ defined in such a way that it possesses
many zeroes, maxima and minima (see Fig. 5). This system describes an array
of inhomogeneities. The array can be studied as a series of elements with
two zeroes and a maximum. If for each element the condition $F_0>\frac{F_mS^2%
}6$ is satisfied, then the soliton can cross the whole inhomogeneous zone
(we have checked this numerically). The array can be completely disordered.
If the condition is fulfilled, there is no localization.

In many systems \cite{Gredeskul,Schrieffer,Davidov,Rice,YuLu} the solitons
play the role of means of transport: they can carry energy and/or charge.
Our result shows that, when soliton tunneling is possible, the soliton can
be a very efficient carrier.

K\"albermann \cite{Kalbermann} investigated impurities that are introduced
in the Hamiltonian density in the following way:

\begin{equation}
H=\frac 12\left( \frac{\partial \phi }{\partial t}\right) ^2+\frac 12\left( 
\frac{\partial \phi }{\partial x}\right) ^2+\frac 18p\left( \phi ^2-1\right)
^2,  \label{Eq6}
\end{equation}
where $p=p_0+U(x)$; $p_0$ is a constant, $U(x)$ is the perturbation that
describes the impurity.

The perturbations used in the numerical experiments in Ref. \cite{Kalbermann}
are given by the function

\begin{equation}
U(x)=\frac{h_1}{\cosh ^2\left( \frac{x-x_1}{a_1}\right) }+\frac{h_2}{\cosh
^2\left( \frac{x-x_2}{a_2}\right) }.  \label{Eq7}
\end{equation}

This perturbation is equivalent to an effective potential with a maximum, a
minimum or a combination of both. Out of the inhomogeneous zone, the
potential tends to zero exponentially. Suppose that the soliton moves from
the left with a kinetic energy less than the maximum of the potential
barrier \cite{Kalbermann}. In this case the soliton tunneling does not
exist. As we have shown, for the soliton tunneling (among other conditions)
it is necessary to have a soliton moving in a potential well where $V(x)$
takes values (out of the potential well) greater than that of the barrier.
Nevertheless, with localized impurities like the ones introduced in Eq. (\ref
{Eq6}), the soliton tunneling can also be observed. This is possible with a
perturbation $U(x)$ with the features shown in Fig. 6. Even in this case, if
the soliton moves from the left (in zone A) with sub-barrier kinetic energy,
then the soliton tunneling does not occur. For the tunneling, the soliton
should be placed in zone B. Of course, the rest of the conditions should be
satisfied.

The sine-Gordon soliton usually is thought to be a very point-like object.
In fact, the unperturbed sine-Gordon equation is integrable and its soliton
solution does not have discrete, internal (shape) modes. Nevertheless, when
the perturbations are not given by Dirac's $\delta $-functions, the
sine-Gordon soliton also is able to excite a great number of shape modes 
\cite{Gonzalez4} and behaves as an extended object.

Consider the perturbed sine-Gordon equation

\begin{equation}
\phi _{tt}-\phi _{xx}+\sin \phi =F(x),  \label{Eq8}
\end{equation}
where

\begin{eqnarray}
F(x)=\cases{-F_0 &for $x<x_0$, \cr \frac a{\cosh ^2(bx)}-F_0^{*} &for
$-x_0\leq x\leq x_0$, \cr -F_0 &for $x>x_0$; }  \label{Eq9}
\end{eqnarray}
$x_0$ and $F_0^{*}$ are chosen such that the function $F(x)$ is continuous.

Let us see only two examples. Let $F_0=0.25$ and $b=0.55$ (fixed). For $%
a=0.45$, the soliton remains trapped in the potential well. It behaves
classically. For $a=0.35$, the soliton escapes from the potential well,
crossing the barrier. We should emphasize that in both cases the force $F(x)$
would correspond to a system with a potential well and a barrier if the
soliton behaves classically.

We would like to remark that, in our study, the positioning of the
soliton at a certain point is done considering the equation of motion. The
initial configuration is always a solution of the equation of motion (we can
use both the static and the time dependent solutions in dependence on the
physical process that led to the given situation). That is, we never use
an initial configuration which the soliton can not reach by any means. For
example, let us explain a physical situation in which the soliton can be
positioned inside the potential well depicted in Fig. 1. Suppose a soliton
is captured by a localized inhomogeneity (there are many such situations
discussed in the review paper \cite{Kivshar1} and the experimental papers
quoted therein). Then, we apply an external constant force (e.g. for the 
Josephson fluxon \cite{Scott} this is a dc bias current). In that case, the 
soliton can be placed in an effective potential similar to that shown in 
Fig. 1. If it had lost all its kinetic energy \cite{Akoh,Kivshar1},
then we can use the static soliton as an initial condition. In fact, we
think that the conditions for soliton tunneling could have been
satisfied in the experimental situation described in \cite{Akoh}. 

We conclude that the phenomenon of soliton tunneling is robust and generic.
We believe this phenomenon can be observed also in other physical systems
bearing solitons, topological defects, vortices, spiral waves, etc.

\begin{figure}[tbp]
\caption{Potential $V(x)$ for the soliton escaping problem. The inset shows
the force $F(x)$. Note that in all figures the quantities plotted
are dimensionless.}
\end{figure}

\begin{figure}[tbp]
\caption{Numerical simulation of the soliton tunneling with sub-barrier
kinetic energy. The pale curve is the potential and the bold curve is the
soliton. The inflexion point is approximately the center of mass. (a) $t=0$, $V(t=0)=0$; (b-f) show the dynamics in succesive time instants.}
\end{figure}

\begin{figure}[tbp]
\caption{Conditions for the existence of soliton tunneling with sub-barrier
kinetic energy. The curve $F_0=\frac{F_mS^2}6$ separates the upper zone,
where the soliton tunneling is possible from the lower zone, where the
soliton tunneling is impossible. The filled circles represent numerical experiments.}
\end{figure}

\begin{figure}[tbp]
\caption{Conditions for the existence of soliton tunneling with sub-barrier
kinetic energy involving the parameters $F_0$ and $V_m$. The other
parameters remain fixed.}
\end{figure}

\begin{figure}[tbp]
\caption{Disordered array of inhomogeneities. The soliton can move through
the whole array.}
\end{figure}

\begin{figure}[tbp]
\caption{Combination of impurities for Eq. (\ref{Eq6}). The soliton
tunneling with sub-barrier kinetic energy is possible from zone B to zone C.}
\end{figure}

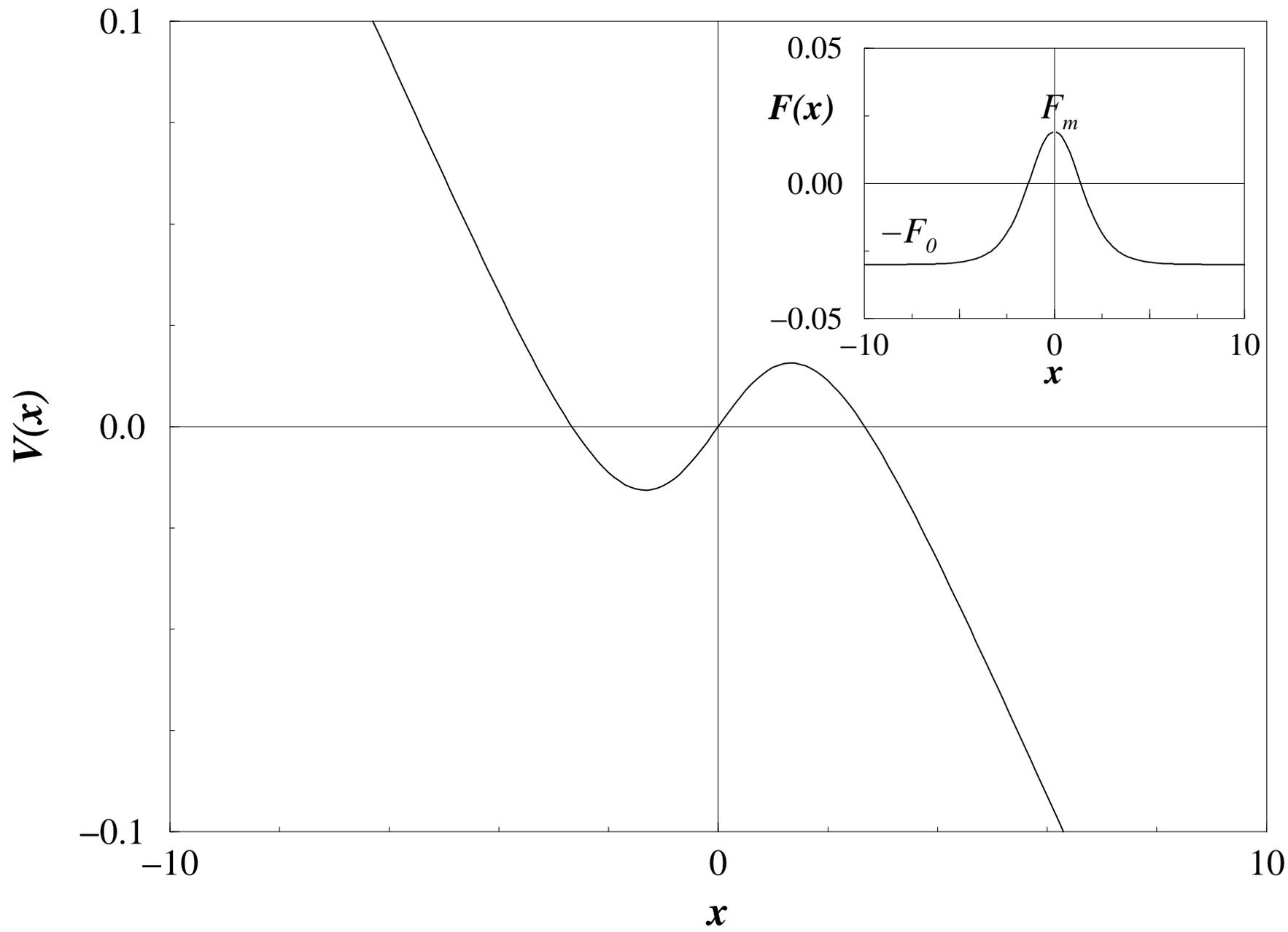
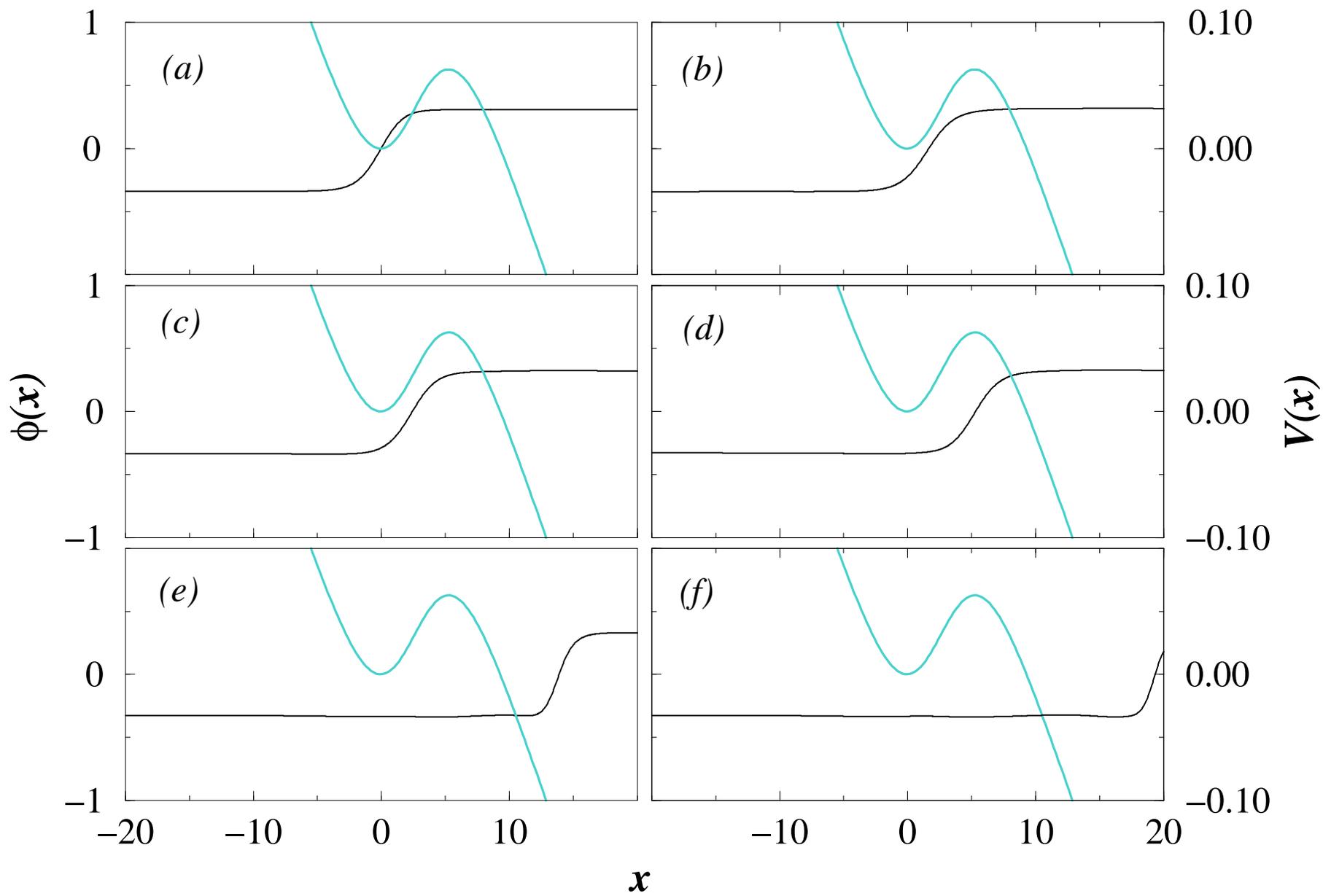
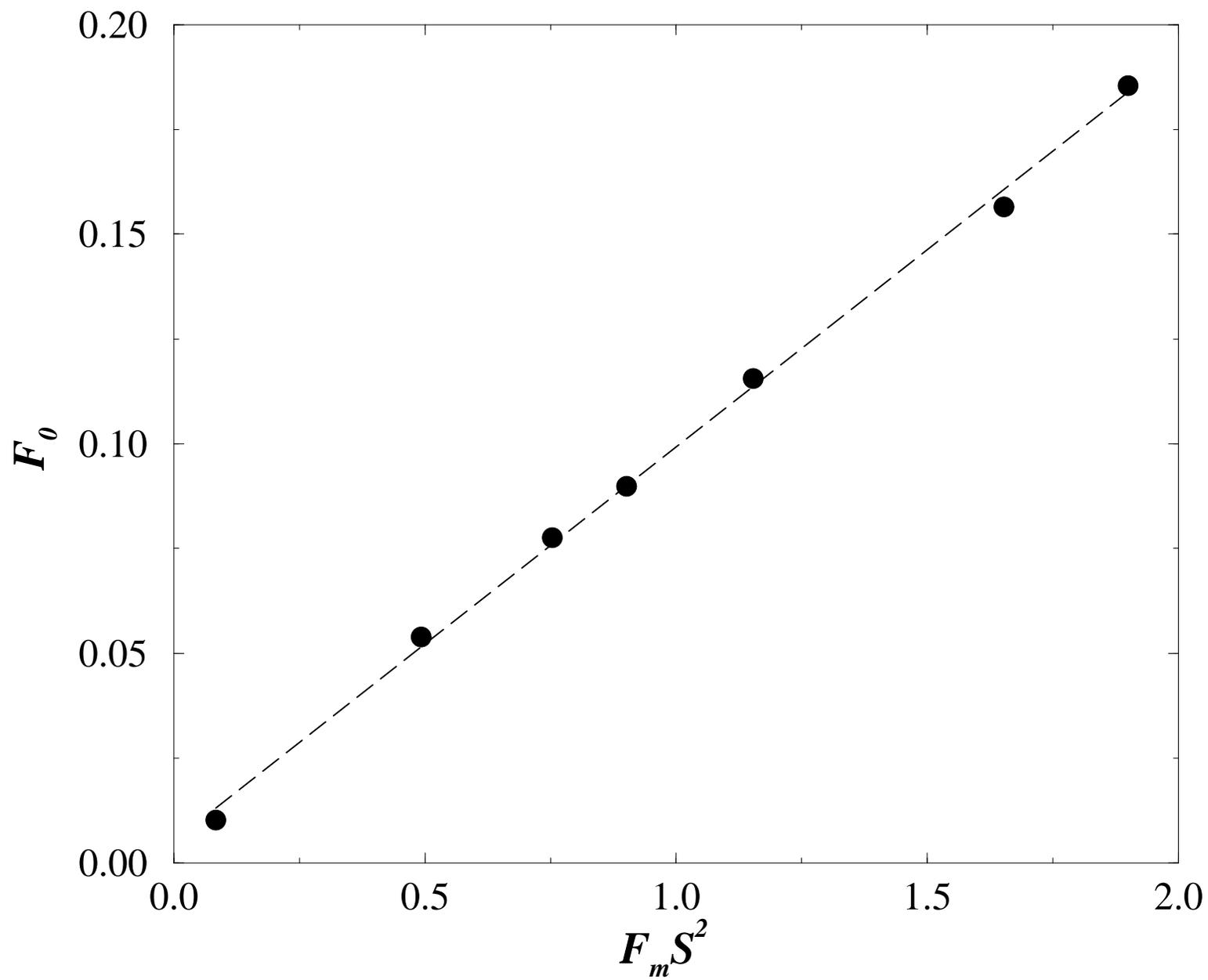
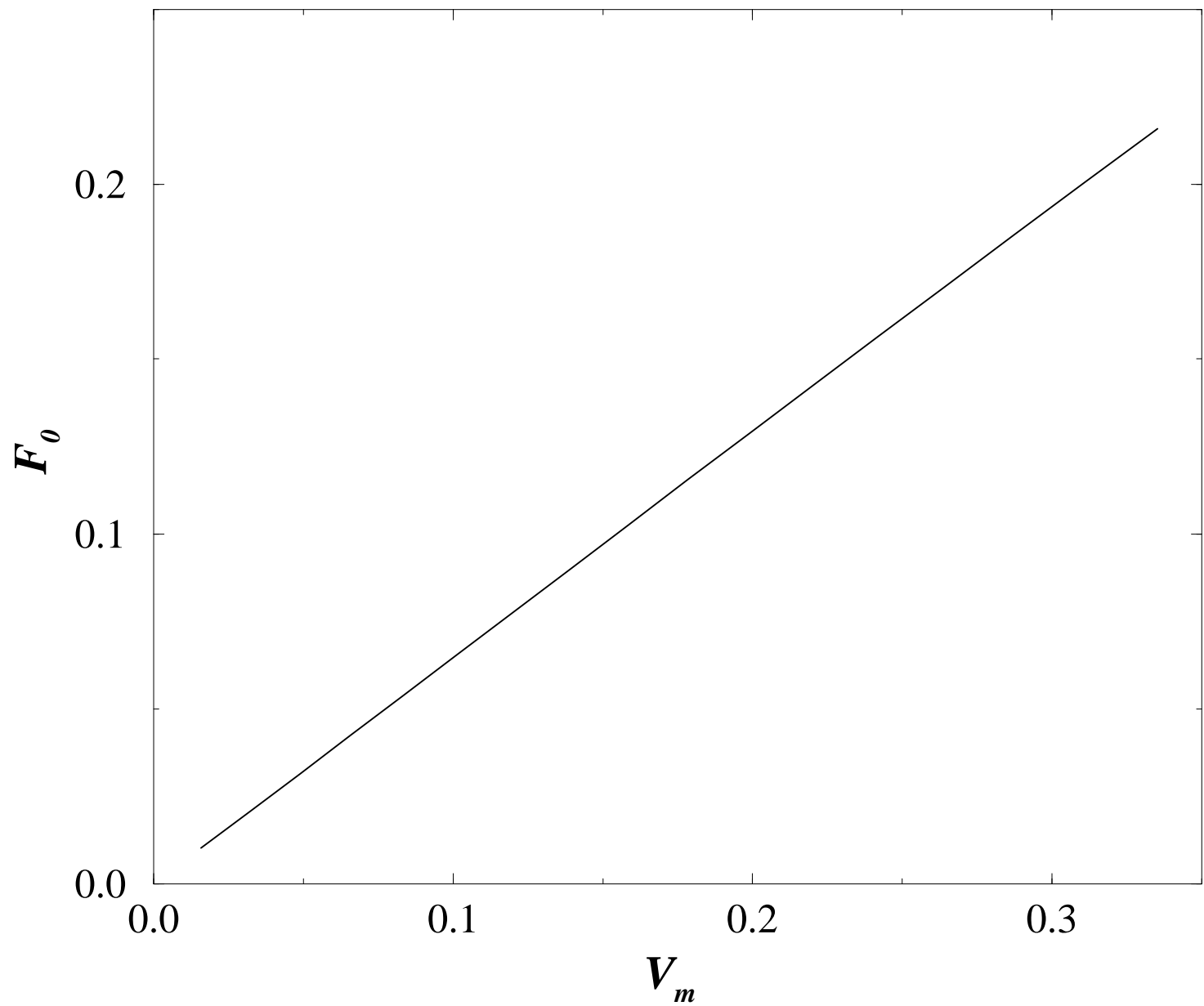
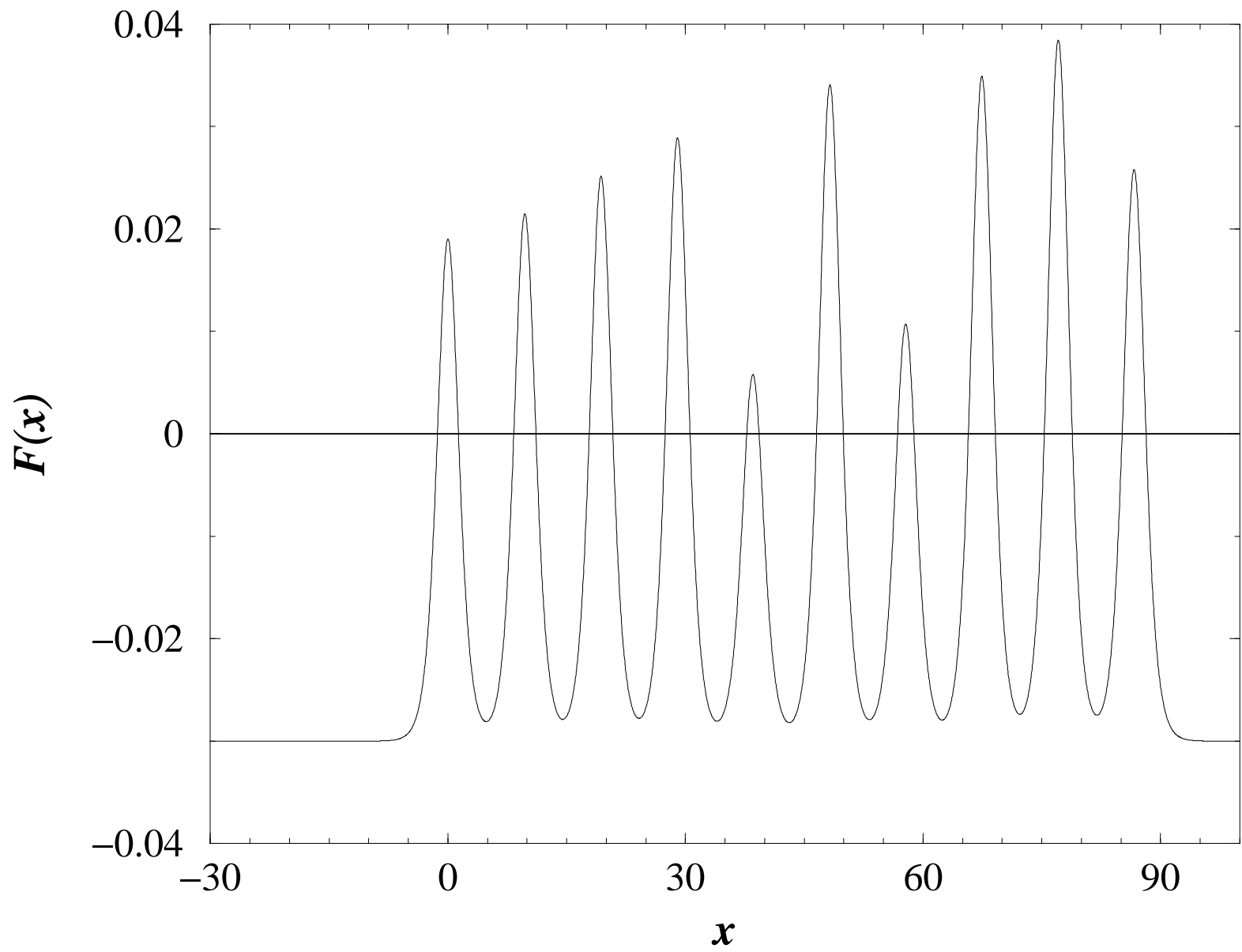
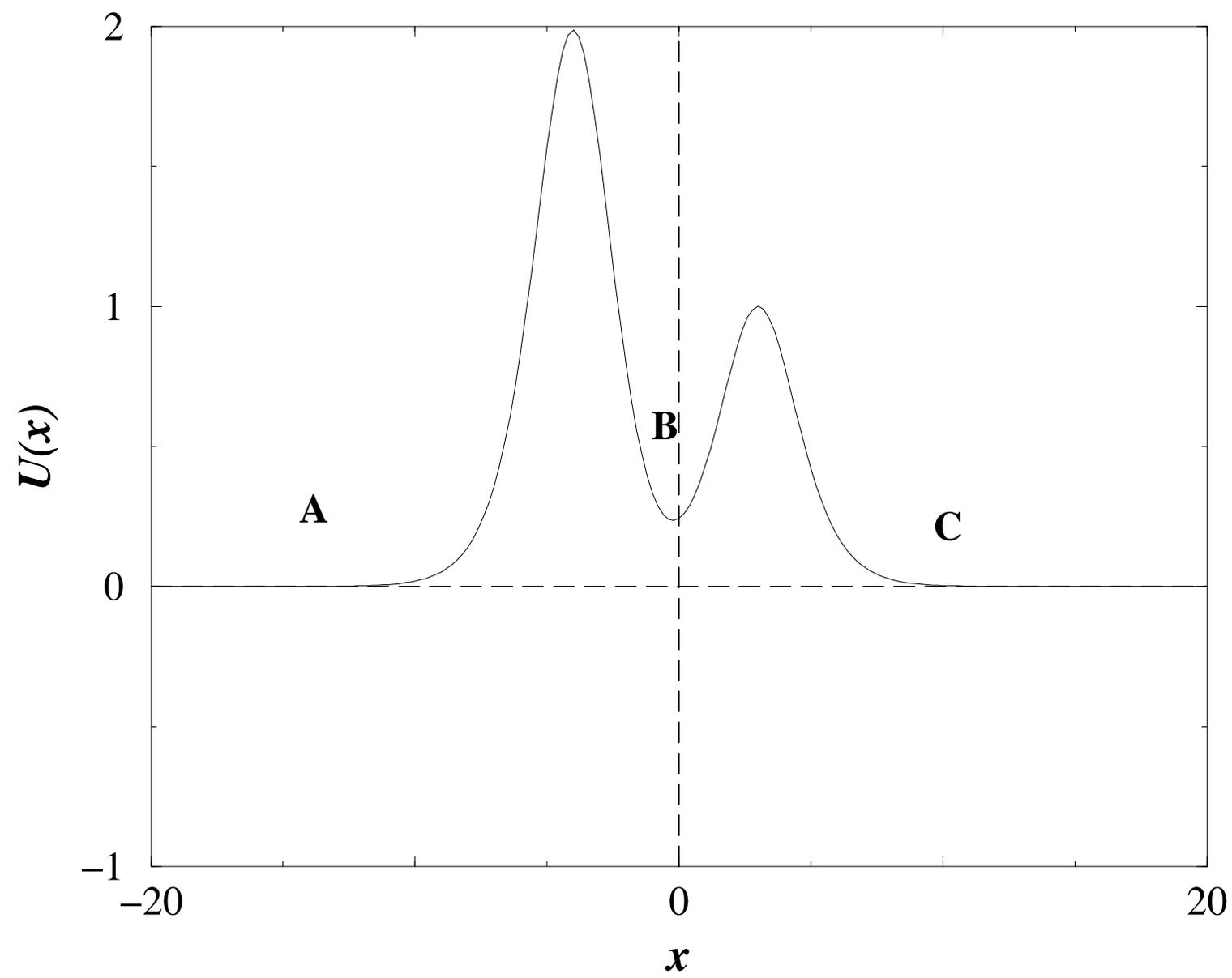
\end{document}